\documentclass[Physsubmission, Phys]{SciPost}

\hypersetup{
    colorlinks,
    linkcolor={red!50!black},
    citecolor={blue!50!black},
    urlcolor={blue!80!black}
}

\usepackage[bitstream-charter]{mathdesign}
\DeclareSymbolFont{usualmathcal}{OMS}{cmsy}{m}{n}
\DeclareSymbolFontAlphabet{\mathcal}{usualmathcal}

\begin{document}

\begin{center}{\Large \textbf{
Highlights from the Telescope Array Experiment\\
}}\end{center}

\begin{center}
Hiroyuki Sagawa\textsuperscript{1$\star$}
\end{center}

\begin{center}
{\bf 1} Institute for Cosmic Ray, the University of Tokyo
\\
for the Telescope Array Collaboration
\\
* hsagawa@icrr.u-tokyo.ac.jp
\end{center}

\begin{center}
\today
\end{center}


\definecolor{palegray}{gray}{0.95}
\begin{center}
\colorbox{palegray}{
  \begin{tabular}{rr}
  \begin{minipage}{0.1\textwidth}
    \includegraphics[width=30mm]{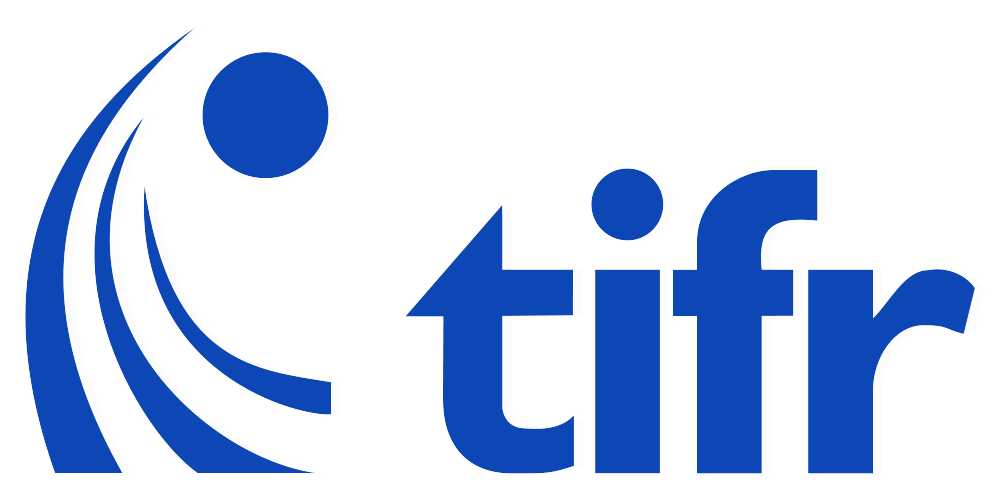}
  \end{minipage}
  &
  \begin{minipage}{0.85\textwidth}
    \begin{center}
    {\it 21st International Symposium on Very High Energy Cosmic Ray Interactions (ISVHE- CRI 2022)}\\
    {\it Online, 23-27 May 2022} \\
    \doi{10.21468/SciPostPhysProc.?}\\
    \end{center}
  \end{minipage}
\end{tabular}
}
\end{center}

\section*{Abstract}
{\bf
The Telescope Array (TA) is the largest hybrid cosmic ray detector in the Northern Hemisphere,  which observes primary particles in the energy range from 2 PeV to 100 EeV. The main TA detector consists of 507 plastic scintillation counters on a 1.2-km spacing square grid and fluorescence detectors at three stations overlooking the sky above the surface detector array. 
The TA Low energy Extension (TALE) detector, which consists of ten fluorescence telescopes, and 80 infill surface detectors with 400m and 600 m spacing, has continued to provide stable observations since its construction completion in 2018. 
The TAx4, a plan to quadruple the detection area of TA is also ongoing. About half of the planned surface detectors have been deployed, and the current TAx4 continues to operate stably as a hybrid detector. I review the present status of the TA experiment and the recent results on the cosmic-ray anisotropy, mass composition and energy spectrum.
}

\vspace{10pt}
\noindent\rule{\textwidth}{1pt}
\tableofcontents\thispagestyle{fancy}
\noindent\rule{\textwidth}{1pt}
\vspace{10pt}

\section{Introduction}
\label{sec:intro}
The Telescope Array (TA) is the largest ultra-high-energy cosmic-ray (UHECR) 
observatory in the northern hemisphere.
The main goal is to explore the origin and nature of UHECRs by
measuring the energy spectrum, arrival direction
distribution and mass composition. 

The TA detector is located in Utah in the U.S.A. and consists of a surface array of 507 plastic scintillator 
detectors (SD)~\cite{bib:SD}, which is overlooked by three stations of 
fluorescence detectors (FD).
The SDs are deployed on a square grid with 1.2-km spacing, and the SD array 
covers an area of approximately 700 km$^2$.
Each individual SD has two layers each of a 1.2-cm-thick scintillator with 
an area of 3 m$^2$.
The full operation of the SDs started in March 2008, and the duty cycle is 
greater than 95\%.
Two FD stations are located at the Black Rock Mesa (BR) and Long Ridge (LR) 
sites~\cite{bib:FD-BRLR}, respectively. At each station, 12 fluorescence 
telescopes, each with 256 photomultipliers (PMTs), cover a total field 
of view of 3$-$31$^\circ$ in elevation angle and 108$^\circ$ in azimuthal angle.
The northern FD station situated at the Middle Drum (MD) site consists of 
14 telescopes refurbished from the HiRes-1 telescopes~\cite{bib:HiRes-1}, 
which were arranged to view $\sim$120$^\circ$ in azimuthal angle.
All three FD stations started the observation in November 2007, and 
they have duty cycles of approximately 10\%.
Hybrid cosmic-ray events, which are detected simultaneously by FD and SD, are 
used to cross-check the SD energy by using the FD energy measurements and to improve mass composition 
identification from longitudinal shower profiles measured with the FD by the inclusion of 
SD information that better determines the directions of air-shower axes.

The TALE enables detailed studies of the energy spectrum and composition 
at energies over $\sim$10$^{16}$ eV.
The main goal of the TALE is to clarify the expected transition from galactic 
to extragalactic cosmic rays.
The TALE FD station is located at the MD site and
consists of 10 telescopes refurbished from HiRes-2~\cite{bib:HiRes-2} and has a field of view of 31$-$59$^\circ$ in elevation angle.
A total of 80 TALE SDs are operating. The data acquisition is ongoing with 
the hybrid trigger from FD. 
Additionally, information on timing around the cores of cosmic-ray air showers 
measured by SDs is expected to improve the event reconstruction accuracy of 
the FD measurement. Consequently, mass composition 
measurements from the longitudinal shower profile with FD are expected to be improved. 
Further low energy extension with hybrid mode is planned with 45 SDs with 100-m spacing and 9 SDs with 200-m spacing. The target energy range is E $>$ 10$^{15}$ eV. The counters were assembled in October 2021.They will be deployed in 2022.

\begin{figure}[!htbp]
\centering 
\includegraphics[width=12cm]{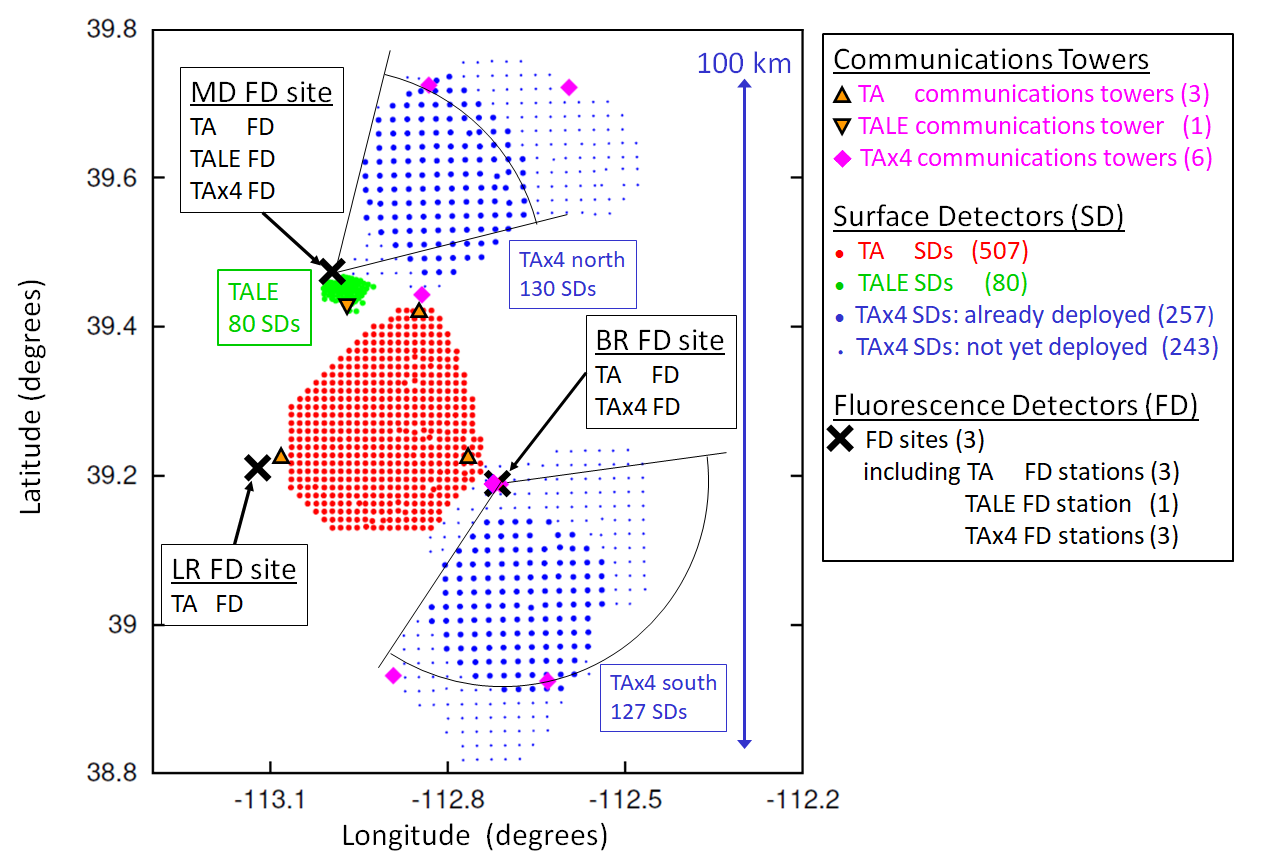}
\caption{\label{fig:TAx4} The layout of TA, TALE and TAx4. The symbols for the layout are explained on the right side of the plot.
Numbers in parentheses indicate the number of detectors.
}
\end{figure}

With enhanced statistics, we expect to verify the hotspot~\cite{bib:TAhotspot-5yrs} along with other 
anisotropy results.
We intended to quadruple the TA aperture (TAx4), including the TA SD array by 
installing 500 SDs at 2.08-km spacing~\cite{ref:TAx4b}.
The construction started in 2015 by reviewing the 
TA scintillator detector components.
A total of 257 SDs were deployed in February and March 2019~\cite{ref:TAx4c}.
The array is 2.5 times larger than the TA SD array. The additional array 
started stable data acquisition in November 2019. 
The new array requires two FD stations overlooking the SD array to increase 
the number of detected hybrid events and to calibrate the energy measured by 
SD. These FDs are formed using refurbished HiRes telescopes.
The first light was observed by the FDs at the northern site and southern site 
in February 2018 and October 2019, respectively. The layout of the 
TAx4 SD and FD including TA and TALE is shown in Fig.~\ref{fig:TAx4}.

In this report, the TA results on spectrum, composition and anisotropy are described in Section~\ref{sec:spectrum}, Section~\ref{sec:composition} and Section~\ref{sec:anisotropy}, respectively.
Section~\ref{sec:conclusion} concludes this report~\cite{bib:JINST-proc-hs}.

\section{Energy Spectrum}
\label{sec:spectrum}
\begin{figure}[!htbp]
\begin{center}
\includegraphics[width=9.1cm]{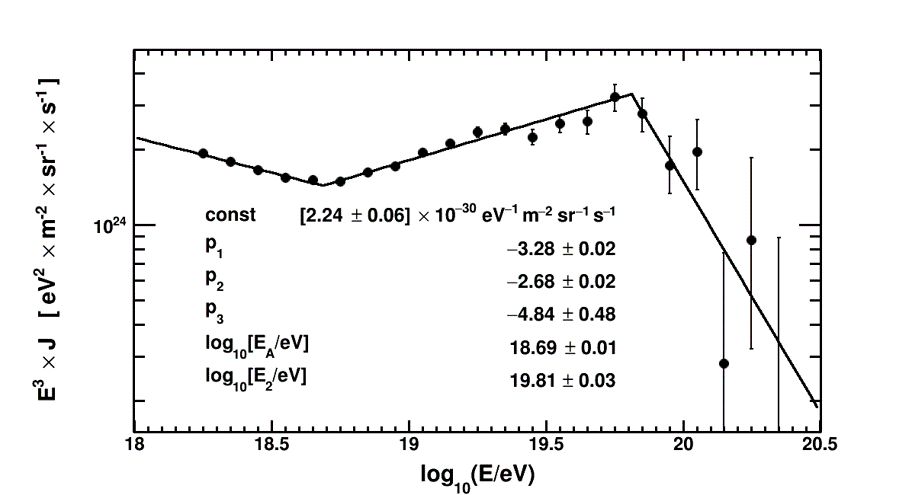}
\includegraphics[width=4.5cm]{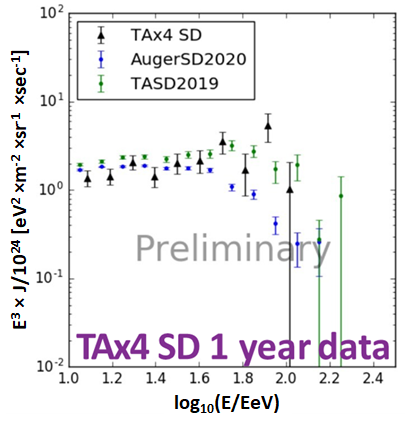}
\end{center}
\vspace{-5mm}
\caption{
(Left) The TA cosmic-ray flux multiplied by E$^3$ measured using 
11 years of data from the TA SD~\cite{bib:icrc2019-TA-DI}.
The solid line shows the fit of the TA data to a broken power law.
(Right) The preliminary spectrum using the first one year of the TAx4 SD data~\cite{bib:icrc2021-TA-HJ} in black together with the Auger spectrum~\cite{bib:Augerspectrum2020} in blue and the TA SD spectrum~\cite{bib:icrc2019-TA-DI} in green.
}
\label{fig:TA-SPECTRUM}
\end{figure}

A preliminary 
result of the cosmic-ray spectrum for 11 years of TA SD data is shown above 10$^{18.2}$ eV in 
Fig.~\ref{fig:TA-SPECTRUM}~(left)~\cite{bib:icrc2019-TA-DI}. 
The TA confirmed the ankle at 10$^{18.69\pm{0.01}}$ eV and the flux suppression 
above 10$^{19.81\pm{0.03}}$ eV, which is consistent with the 
Greisen-Zatsepin-Kuzmin cutoff prediction~\cite{bib:GZK1,bib:GZK2}. The statistical 
significance of the difference between the observed cosmic-ray flux above 
the cutoff and the expected one from the extrapolation of the flux with the spectral index between the ankle and 
the cutoff (no suppression) is $\sim$8.4$\sigma$. 

Fig.~\ref{fig:TA-SPECTRUM}~(right)~\cite{bib:icrc2021-TA-HJ} shows the preliminary result of TAx4 SD spectrum, which is
consistent with the TA SD spectrum.
Hereafter, when referring to TAx4 results, it means results using only data from 257 SDs deployed for TAx4 in 2019.
The monocular energy spectrum using the additional TAx4 FD~\cite{bib:icrc2021-TA-MP} and 
the energy spectrum using the hybrid analysis with TAx4~\cite{bib:aps-s2022-TA-MP} are also consistent with the TA SD spectrum.

The energy spectra measured by the Pierre Auger Observatory (hereafter Auger)~\cite{bib:Augerspectrum2020} in the southern 
hemisphere and by the TA~\cite{bib:icrc2019-TA-DI} in the northern hemisphere agree well for energies
(0.1$-$2.5) $\times$ 10$^{19}$ eV 
after rescaling the energies by +4.5\% for Auger and $-$4.5\% for TA,
whereas there is a significant discrepancy between the two results 
at the suppression~\cite{bib:icrc2019-TA-DI}. 
When we compare the energy spectra in the common declination band ($-$15$^\circ$ $<$ $\delta$ $<$ +24.8$^\circ$) after the $\pm$4.5\% rescalings, the differences are smaller, but the persistent differences require an additional energy rescaling in an energy-dependent way ($\pm$10\%/decade for E $>$ 10$^{19}$ eV) to get an agreement~\cite{ref:TA-Auger-spectrumWG-icrc2021}.

\begin{figure}[!htbp]
\begin{center}
\includegraphics[width=7.5cm]{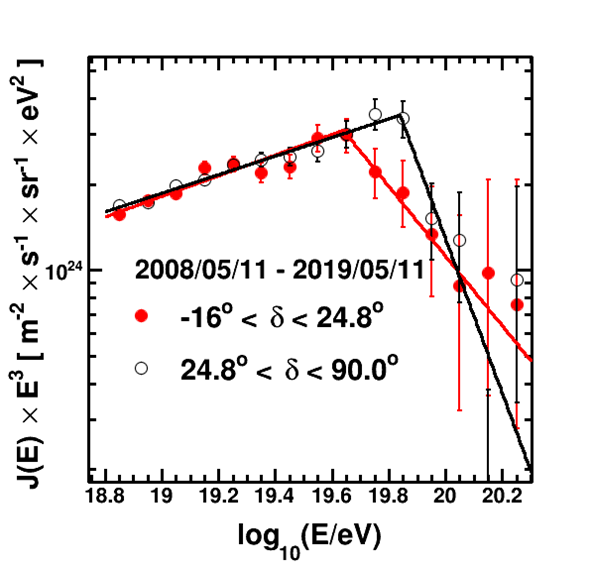}
\end{center}
\vspace{-5mm}
\caption{
The TA cosmic-ray flux multiplied by E$^3$ measured using 11 years of data 
obtained by the TA SD for the upper (24.8$^\circ$ $<$ $\delta$ $<$ $90^\circ$) 
and lower ($-$16$^\circ$ $<$ $\delta$ $<$ 24.8$^\circ$) declination bands with 
black open circles and red closed circles, respectively~\cite{bib:icrc2019-TA-DI}. Superimposed lines 
correspond to the fits of the data to broken power-law functions. 
}
\label{fig:TAspectrum-decl}
\end{figure}

As shown in Fig~\ref{fig:TAspectrum-decl}, the TA SD data over 11 years 
yield the cutoffs at 10$^{19.64\pm{0.04}}$ eV and 10$^{19.84\pm{0.02}}$ eV for  
declinations of $-$16$^\circ$ $<$ $\delta$ $<$ 24.8$^\circ$ and 24.8$^\circ$ $<$ $\delta$ $<$ 90$^\circ$, respectively. 
The pretrial significance of the difference of these two cutoffs is 
4.7$\sigma$. The chance probability of exceeding the pretrial significance 
for an isotropic distribution is 8.5$\times$10$^{-6}$ or 4.3$\sigma$~\cite{bib:icrc2019-TA-DI}
%

\begin{figure}[!htbp]
\begin{center}
\includegraphics[width=8.5cm]{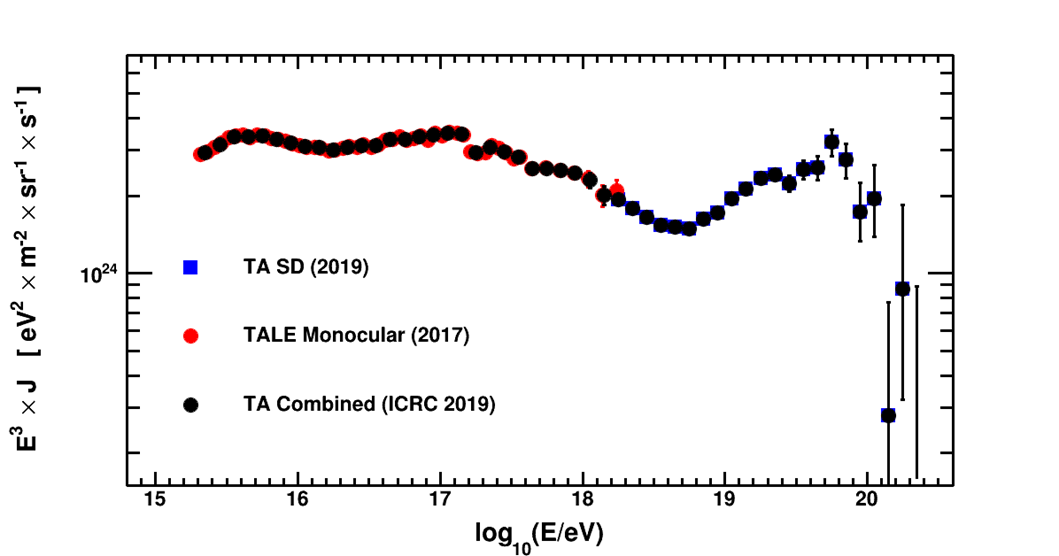}
\includegraphics[width=6cm]{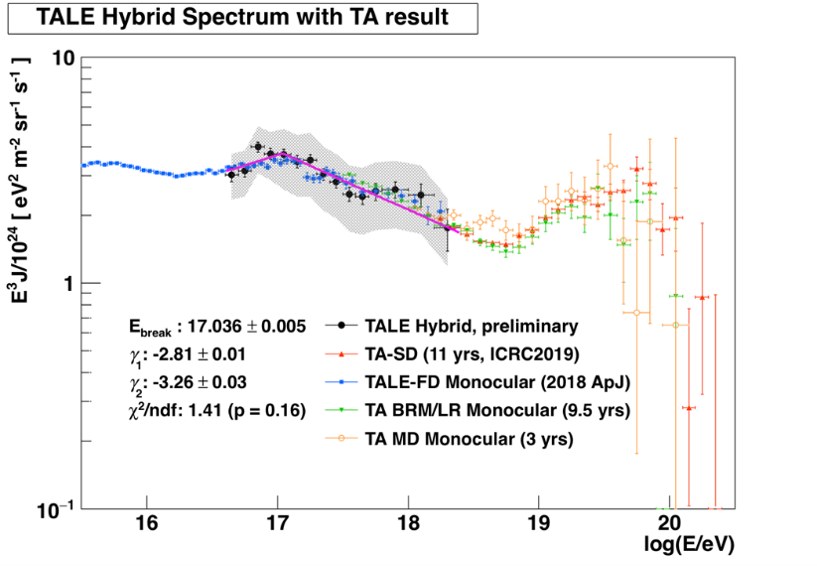}
\end{center}
\vspace{-5mm}
\caption{
(Left) The TA combined spectrum in black, which is made by combining the TALE FD monocular data in red 
and the TA SD data in blue~\cite{bib:TAspectrum2021}.
(Right) The preliminary TALE hybrid spectrum in black together with other measurements from TA and TALE~\cite{bib:TALEhybridspectrum-isvhecri2022}.
}
\label{fig:TA-TALE-spectrum}
\end{figure}

The TALE FD measures a spectrum with mixed Cherenkov and fluorescence signals.
The energy spectrum collected by the TALE FD monocular measurement over 22 months was published in~\cite{bib:TALEspectrum}.
Fig.~\ref{fig:TA-TALE-spectrum} (left)~\cite{bib:TAspectrum2021} shows the TA combined spectrum, which is made by combining the TA SD and TALE FD spectra. 
The range of energies below 10$^{18.2}$ eV is covered by the TALE FD, while the energy range above 10$^{18.2}$ eV is covered  by the TA SD.
We see three features in the energy spectrum: the knee at approximately 10$^{15.5}$ eV, the 
low-energy ankle at 10$^{16.22\pm{0.02}}$ eV and the second knee at 
10$^{17.04\pm{0.04}}$ eV.
Fig.~\ref{fig:TA-TALE-spectrum} (right) shows the preliminary spectrum using 2.5 years of TA hybrid data, which is consistent with other TA/TALE spectra within systematic uncertainties~\cite{bib:TALEhybridspectrum-isvhecri2022}.

\begin{figure}[!htbp]
\begin{center}
\includegraphics[width=14cm]{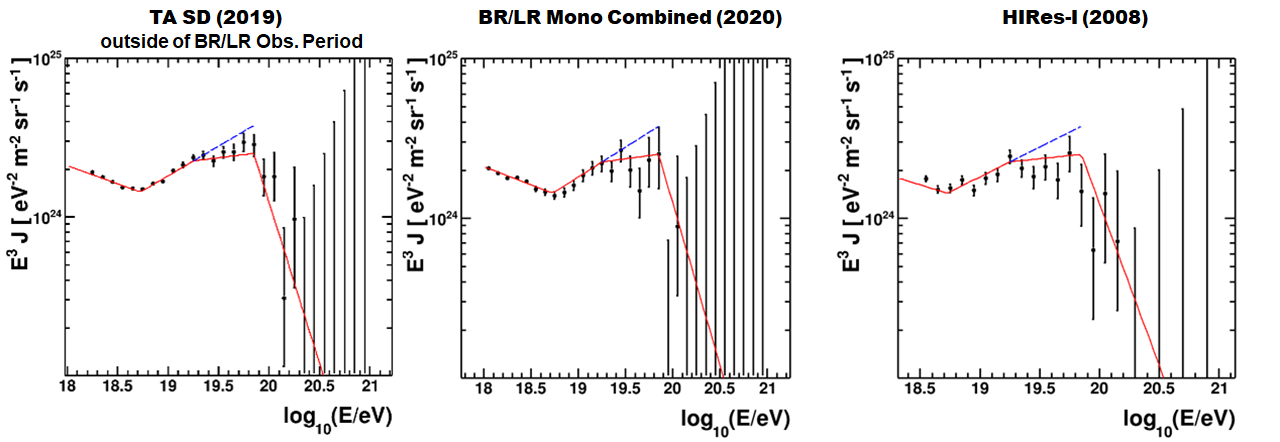}
\end{center}
\vspace{-5mm}
\caption{
A joint fit~\cite{bib:TAspectrum2021} to the TA SD (left), TA BR - LR monocular FD (center), and HiRes I monocular energy spectra (right) together with a red fit line with
three break points.
The significance of the shoulder is obtained by comparing the number of events expected in the absence of the feature (blue line) and the number of events observed by the experiments.
}
\label{fig:TA-spectrum-instep}
\end{figure}

A new feature above 10$^{19}$ eV, called the shoulder or instep, was first reported by Auger, of which field of view is concentrated in the southern sky~\cite{ref:Augerspectrum-instep}. We performed a joint fit of  TA SD, TA Black Rock (BR) - Long Ridge (LR) monocular FD and HiRes I monocular spectra in the northern hemisphere into a thrice broken power law. To ensure statistical independence, the TA monocular FD observation period was removed from the TA SD spectrum measurement. The shoulder feature as shown in Fig.~\ref{fig:TA-spectrum-instep} was found at 10$^{19.25{\pm}0.03}$ eV with a statistical significance of 5.3$\sigma$~\cite{bib:TAspectrum2021}.

\section{Composition}
\label{sec:composition}
\begin{figure}[!htbp]
\begin{center}
\includegraphics[width=7cm]{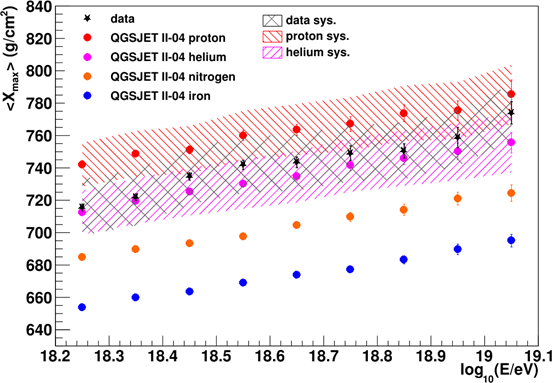}
\includegraphics[width=7cm]{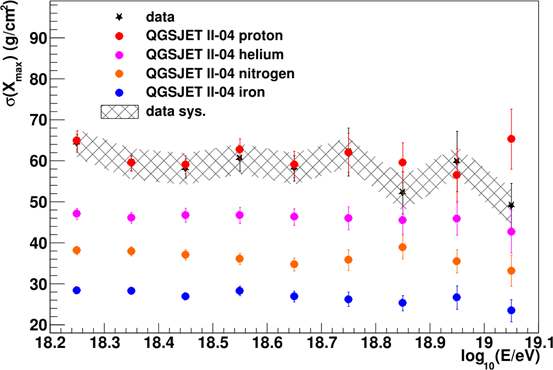}
\end{center}
\vspace{-5mm}
\caption{
The results of the mean $X_{\rm max}$ ($<{X_{\rm max}}>$) (left) and the 
width of $X_{\rm max}$ ($\sigma{(X_{\rm max})}$) (right) as a function of the cosmic-ray 
energy measured from TA hybrid events~\cite{bib:TAcompostion-2019}.
}
\label{fig:TA-Xmax-ave-width}
\end{figure}

The depth of shower maximum, $X_{\rm max}$, of extensive air-shower profiles is the 
key estimate of the mass composition of UHECRs. The results of $X_{\rm max}$ 
measurements in the energy range between 10$^{18.2}$ eV and 10$^{19.1}$ eV 
based on the TA hybrid events observed over 10 years are shown in Fig.~\ref{fig:TA-Xmax-ave-width} together with MC predictions of QGSJET II-04~\cite{bib:QGSJET-II-04} for proton, helium, nitrogen, and iron primaries~\cite{bib:TAcompostion-2019}. 
We need more statistics to clarify the feature of $X_{\rm max}$ above 10$^{19}$ eV, for example, to more accurately measure the values of the mean $X_{\rm max}$ and the width of $X_{\rm max}$.
The model predictions for proton and helium primaries appear to agree with the data within systematic uncertainties.

\begin{figure}[!htbp]
\begin{center}
\includegraphics[width=9cm]{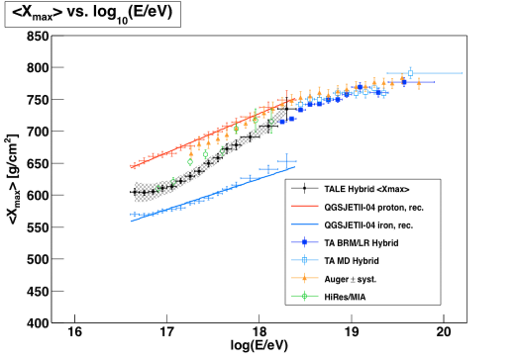}
\end{center}
\vspace{-5mm}
\caption{
The preliminary result of $<X_{\rm max}>$ of the reconstructed TALE hybrid events as a function of 
energy~\cite{bib:TALEhybridspectrum-isvhecri2022}.
The $<X_{\rm max}>$ values for proton and iron MC primary elements, using QGSJET II-04, are also shown. 
}
\label{fig:TALEcomposition}
\end{figure}

The $X_{\rm max}$ above 10$^{15.3}$ $-$ 10$^{18.3}$ eV was measured using the TALE FD data and a change in the elongation rate was observed at an energy of $\sim$10$^{17.3}$ eV~\cite{{bib:TALEcomposition}}. The $X_{\rm max}$ was measured using the TALE hybrid data collected over a period 
of 2.5 years in the energy range 10$^{16.6}$ $-$ 10$^{18.4}$ eV~\cite{bib:TALEhybridspectrum-isvhecri2022}.
Figure~\ref{fig:TALEcomposition} shows the mean $X_{\rm max}$ as a function 
of the cosmic-ray energy.
The systematic uncertainty on $X_{\rm max}$ is less than 16 g/cm$^2$.
The result is not corrected for a bias of about $-$10 g/cm$^2$.
We can see a break point at $\sim$10$^{17}$ eV, which is likely correlated with the observed softening 
($\sim$10$^{17}$ eV) in the TALE energy spectrum~\cite{bib:TALEspectrum}.
The values of the mean $X_{\rm max}$ and the slope of the $X_{\rm max}$ elongation ratio for the TALE hybrid analysis are consistent with those for the TA hybrid analysis at around 10$^{18.2}$ eV.

The composition information was also derived by the Boosted Decision Trees (BDT) method using 16 composition sensitive signals from 12 years of TA SD data~\cite{ref:SDcomposition-icrc2021}. The measured mean logarithmic cosmic-ray atomic number $<ln A>$ shows no significant energy dependence above 1 EeV and $<ln A>$ = 0.90 $\pm$ 0.05(stat.) $\pm$ 0.30(syst.) with the QGSJET II-04 model.

\section{Anisotropy}
\label{sec:anisotropy}
\begin{figure}[!htbp]
\begin{center}
\includegraphics[width=7.5cm]{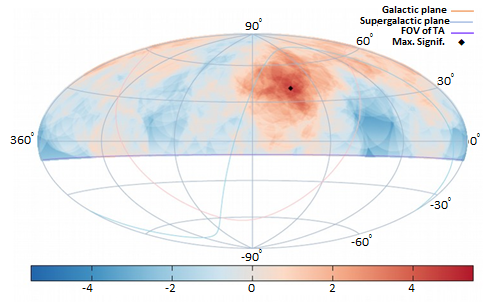}
\end{center}
\vspace{-5mm}
\caption{
The statistical significance of UHECR events with $E >$ 5.7$\times$10$^{19}$ eV
in equatorial coordinates~\cite{bib:TAhotspot-12yrs}. 
The TA events observed over 12 years are smoothed by the 25$^\circ$ 
oversampling radius circle. 
The solid red and blue curves indicate the galactic (GP) and supergalactic (SGP) planes, respectively.
}
\label{fig:TAhotspot}
\end{figure}

The TA has previously reported an indication of an intermediate-scale cluster 
of cosmic rays with energies over 5.7$\times$10$^{19}$ eV for five years of 
observation with the TA SD\cite{bib:TAhotspot-5yrs}. This TA hotspot result 
has since been updated using 12 years of data from the TA 
SD~\cite{bib:TAhotspot-12yrs}. Figure~\ref{fig:TAhotspot} shows the 
significance maps of 179 UHECR events with $E >$ 5.7$\times$10$^{19}$ eV.
We found the maximum significance of 5.1$\sigma$ at R.A.=144.0$^\circ$ and Dec.=40.5$^\circ$ for the oversampling circle with a radius of 25$^\circ$ after searching for the maximum significance in circles with all grid directions and five different oversampling radii  
(40 events were observed where 14.6 would be expected).
The chance probability of the 12-year hotspot in an isotropic sky is 
estimated to be 6.8$\times$10$^{-4}$ or 3.2$\sigma$. 

\begin{figure}[!htbp]
\begin{center}
\includegraphics[width=7.5cm]{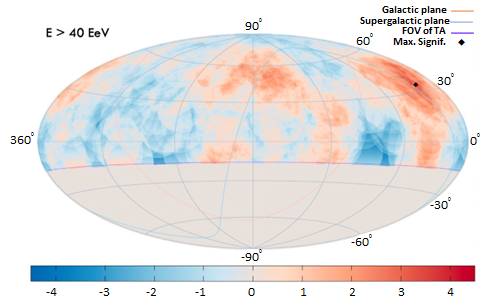}
\end{center}
\vspace{-5mm}
\caption{
The statistical significance of UHECR events with $E >$ 4$\times$10$^{19}$ eV
in equatorial coordinates~\cite{bib:TAhotspot-12yrs}. 
The TA events observed over 11 years are smoothed by the 20$^\circ$ 
oversampling radius circle. 
The solid red and blue curves indicate the galactic (GP) and supergalactic (SGP) planes, respectively.
}
\label{fig:TANewExcess}
\end{figure}

Now we lower the energy threshold slightly. The significances of the TA 11-year data at energies log$_{10}$(E/eV) $>$ 19.4, 19.5, and 19.6 by the oversampling analysis with 20$^\circ$-radius circle are 4.4$\sigma$, 4.2$\sigma$, and 4$\sigma$, respectively (see Fig.~\ref{fig:TANewExcess}). A new excess was found in the direction of the Perseus-Pisces Supercluster. For E $>$ 10$^{19.6}$ eV, the chance probability of the excess within 6.8$^\circ$ from the supercluster center is 3.5$\sigma$.

\section{Conclusion}
\label{sec:conclusion}
TA confirmed the ankle at 10$^{18.69}$ eV and the flux suppression above 
10$^{19.81}$ eV in the energy spectrum of UHECRs. 
The shoulder feature in the energy spectrum in the northern sky was found at 10$^{19.25}$ eV.
We confirmed the breaks at 10$^{16.22}$ eV and 10$^{17.04}$ eV in the energy 
spectrum measured using the data observed with the TALE FD.

The TA hybrid $X_{\rm max}$ measurements 
for the energy band 
10$^{18.2}$$-$10$^{19.1}$ eV are consistent with a light composition, in particular, with predictions of QGSJET II-04 proton and helium. 
We need more statistics to clarify the feature of $X_{\rm max}$ above 10$^{19}$ eV.
For the TALE FD $X_{\rm max}$ results in the energy range 10$^{15.3}$ $-$ 10$^{18.3}$ eV, a change in the elongation rate was observed at an energy 
of $\sim$10$^{17.3}$ eV, which is likely correlated with the observed break 
at 10$^{17.04}$ eV in the TALE FD energy spectrum.

We obtained 179 events above 5.7$\times$10$^{19}$ eV in 12 years 
of observation with the TA SD. 
We found a maximum pretrial significance of 5.1$\sigma$ when using a circle with a 25$^\circ$ oversampling radius. 
The post-significance of detecting such clustered events by chance in the isotropic arrival distribution is estimated to be 3.2$\sigma$.
Evidences for some features of anisotropy are found, for example, as a declination dependence of spectrum cutoff in the energy spectrum, and a new excess in the direction of Perseus-Pisces Supercluster using a slightly lower energy threshold.

To confirm the TA hotspot and understand its features, we proposed a plan, which we call the TAx4, 
to quadruple the TA SD aperture and add two FD stations.
The 257 TAx4 SDs were deployed in 2019, and currently the total
area is 2.5 times the TA SD. 
Two TAx4 FD stations are operating.

The 80 TALE SDs were deployed at the TALE site.
The preliminary result of the spectrum and $X_{\rm max}$ using the TALE hybrid data was present in this conference.
The TALE infill SD array is planned to further lower the energy threshold for the TALE hybrid analysis.

The TA, TAx4, and TALE will provide important measurements of the energy 
spectrum, composition, and arrival directions of UHECRs from the knee region 
up to the highest-energy region spanning five to six decades in energy.

\section*{Acknowledgements}

The Telescope Array experiment is supported by the Japan Society for
the Promotion of Science (JSPS) through
Grants-in-Aid
for Priority Area
431,
for Specially Promoted Research
JP21000002,
for Scientific  Research (S)
JP19104006,
for Specially Promoted Research
JP15H05693,
for Scientific  Research (S)
JP19H05607,
for Scientific  Research (S)
JP15H05741,
for Science Research (A)
JP18H03705,
for Young Scientists (A)
JPH26707011,
and for Fostering Joint International Research (B)
JP19KK0074,
by the joint research program of the Institute for Cosmic Ray Research (ICRR), The University of Tokyo;
by the Pioneering Program of RIKEN for the Evolution of Matter in the Universe (r-EMU);
by the U.S. National Science
Foundation awards PHY-1607727, PHY-1712517, PHY-1806797, PHY-2012934, and PHY-2112904;
by the National Research Foundation of Korea
(2017K1A4A3015188, 2020R1A2C1008230, \& 2020R1A2C2102800) ;
by the Ministry of Science and Higher Education of the Russian Federation under the contract 075-15-2020-778, IISN project No. 4.4501.18, and Belgian Science Policy under IUAP VII/37 (ULB). This work was partially supported by the grants ofThe joint research program of the Institute for Space-Earth Environmental Research, Nagoya University and Inter-University Research Program of the Institute for Cosmic Ray Research of University of Tokyo. The foundations of Dr. Ezekiel R. and Edna Wattis Dumke, Willard L. Eccles, and George S. and Dolores Dor\'e Eccles all helped with generous donations. The State of Utah supported the project through its Economic Development Board, and the University of Utah through the Office of the Vice President for Research. The experimental site became available through the cooperation of the Utah School and Institutional Trust Lands Administration (SITLA), U.S. Bureau of Land Management (BLM), and the U.S. Air Force. We appreciate the assistance of the State of Utah and Fillmore offices of the BLM in crafting the Plan of Development for the site.  Patrick A.~Shea assisted the collaboration with valuable advice and supported the collaboration's efforts. The people and the officials of Millard County, Utah have been a source of steadfast and warm support for our work which we greatly appreciate. We are indebted to the Millard County Road Department for their efforts to maintain and clear the roads which get us to our sites. We gratefully acknowledge the contribution from the technical staffs of our home institutions. An allocation of computer time from the Center for High Performance Computing at the University of Utah is gratefully acknowledged.


\nolinenumbers


\begin{thebibliography}{99}

\bibitem{bib:SD}
T. Abu-Zayyad, et al., \emph{The surface detector array of the Telescope Array experiment}, Nuclear Instruments and Methods in Physics Research Section A {\bf 689} (2012) 87-97.

\bibitem{bib:FD-BRLR}
H. Tokuno, et al., \emph{On site calibration for new fluorescence detectors of the Telescope Array experiment}, Nuclear Instruments and Methods in Physics Research Section A {\bf 601} (2009) 364-371; H. Tokuno, et al., \emph{New air fluorescence detectors employed in the Telescope Array experiment}, Nuclear Instruments and Methods in Physics Research Section A {\bf 676} (2012) 54-65.

\bibitem{bib:HiRes-1}
R.U. Abbasi, et al., \emph{Measurement of the flux of ultrahigh energy cosmic rays from monocular observations 
by the High Resolution Fly's Eye experiment}, Physical Review Letters {\bf 92} (2004) 151101 (4pp).

\bibitem{bib:HiRes-2}
J.H.~Boyer, B.C.~Knapp, E.J.~Mannel, M.~Seman, \emph{FADC-based DAQ for HiRes Fly's Eye}, Nuclear Instruments and Methods in Physics Research Section A {\bf 482} (2002), 457-474.

\bibitem{bib:TAhotspot-5yrs}
R.U.~Abbasi, et al., \emph{Indication of intermediate-scale anisotropy of cosmic rays with energy greater than  57 EeV in the northern sky measured with the surface detector of the Telescope Array experiment}, The Astrophysical Journal Letters, {\bf 790} (2014) L21 (5pp). 

\bibitem{ref:TAx4b}
E. Kido, et al., \emph{The TAx4 experiment}, in proceedings of 35th International Cosmic Ray Conference (ICRC2017), PoS(ICRC2017)386 (2018) (8pp). 

\bibitem{ref:TAx4c}
E. Kido, et al., \emph{Status and prospects of the TAx4 experiment}, in proceedings of 36th International Cosmic Ray Conference (ICRC2019), PoS(ICRC2019)312 (2020) (5pp). 

\bibitem{bib:JINST-proc-hs}
Some preliminary results can be found in proceedings such as H. Sagawa, et al., \emph{Results of ultra-high-energy cosmic rays from the Telescope Array}, Journal of Instrumentation {\bf 15}  (2022) C09012 (10pp). 

\bibitem{bib:icrc2019-TA-DI}
D.~Ivanov, et al., \emph{Energy spectrum measured by the Telescope Array experiment}, 
PoS(ICRC2019)298 (2020) (7pp).

\bibitem{bib:GZK1}
K.~Greisen, \emph{End to the cosmic-ray spectrum?}, Physical Review Letters {\bf 16} (1966) 748-750.

\bibitem{bib:GZK2}
G.T.~Zatsepin and V.A.~Kuz'min, \emph{Upper limit of the spectrum of cosmic rays}, Soviet Physics Journal of Experimental and Theoretical Physics (JETP) Letters {\bf 4} (1966) 114-117.

\bibitem{bib:icrc2021-TA-HJ}
H.~Jeong, et al., 
\emph{Reconstruction of air shower events measured by the surface detectors of the TAx4 experiment}, 
 Pos(ICRC2021)331 (2022) (10pp).
 
\bibitem{bib:icrc2021-TA-MP}
M.~Potts, et al., 
\emph{Monocular energy spectrum using the TAx4 fluorescence detector}, 
 Pos(ICRC2021)343 (2022) (11pp).

\bibitem{bib:aps-s2022-TA-MP}
M.~Potts, et al., 
 \emph{Ultra-high energy cosmic ray energy spectrum using hybrid analysis with TAx4}, 
 American Physical Society April Meeting (2022).
 
\bibitem{bib:Augerspectrum2020}
A.~Aab, et al. (Pierre Auger Collaboration), 
\emph{Measurement of the cosmic-ray energy spectrum above 2.5 $\times$ 10$^{18}$ eV using the Pierre Auger Observatory},
Physical Review D {\bf 102}, 062005 (2020) (27pp).

\bibitem{ref:TA-Auger-spectrumWG-icrc2021}
Y. Tsunesada, for the Telescope Array and Pierre Auger Collaborations, 
\emph{Joint analysis of the energy spectrum of ultra-high-energy cosmic rays measured at the Pierre Auger Observatory and the Telescope Array}, PoS(ICRC2021)337 (2022) (14pp).

\bibitem{bib:TAspectrum2021}
D.~Ivanov, et al., \emph{Recent measurement of the Telescope Array energy spectrum and observation of the shoulder feature in the northern hemisphere}, PoS(ICRC2021)341 (2022) (7pp).

\bibitem{bib:TALEspectrum}
R.U.~Abbasi, et al., \emph{The Cosmic ray energy spectrum between 2~PeV and 2~EeV observed with the TALE detector in monocular mode}, The Astrophysical Journal, {\bf 865} (2018) 74 (18pp).

\bibitem{bib:TALEhybridspectrum-isvhecri2022}
K. Fujita et al.,
\emph{The Telescope Array Low-energy Extension hybryd detector}
in  proceedings of 21st International Symposium on Very High Energy Cosmic Ray Interactions (ISVHECRI2022).

\bibitem{ref:Augerspectrum-instep}
A. Aab, et al. (Pierre Auger Collaboration),
\emph{Features of the energy spectrum of cosmic rays above 2.5 $\times$ 10$^{18}$ eV using the Pierre Auger Observatory}, Phys. Rev. Lett. {\bf 125} (2020) no. 12, 121106 (10pp).

\bibitem{bib:QGSJET-II-04}
S.~Ostapchenko, \emph{Monte Carlo treatment of hadronic interactions in enhanced Pomeron scheme: QGSJET-II model}, Physical Review D, {\bf 83} (2011) 014018 (27pp).

\bibitem{bib:TAcompostion-2019}
W.~Hanlon, et al., \emph{Telescope Array 10 year composition}, PoS(ICRC2019)280 (2020) (7pp).

\bibitem{bib:TALEcomposition}
T.~AbuZayyad, et al., \emph{TALE FD cosmic rays composition measurement}, 
PoS(ICRC2019)169 (2020) (5pp).


\bibitem{ref:SDcomposition-icrc2021}
Y.~Zhezher, et al., \emph{Cosmic-ray mass composition with the TA SD 12-year data}, 
PoS(ICRC2021)300 (2022) (11pp).

\bibitem{bib:TAhotspot-12yrs}
J.~Kim, et al., \emph{Hotspot update, and a new excess of events on the sky seen by the  Telescope Array experiment}, 
PoS(ICRC2021)328 (2022) (10pp).
\end{thebibliography}
\end{document}